\documentclass[aps,prl,twocolumn,showpacs,superscriptaddress]{revtex4}  
\usepackage{graphicx}  
\usepackage{dcolumn}   
\usepackage{bm}        
\usepackage{amssymb}   
\usepackage{multirow}
\usepackage{amsmath}

\begin{document}


\title{$\pi$-Conjugation in Epitaxial Si(111)-($\sqrt{3}$$\times$$\sqrt{3}$) Surface: an Unconventional\\ ``Bamboo Hat" Bonding Geometry for Si}

\author{Wei Jiang}
\affiliation{Department of Materials Science and Engineering, University of Utah, Salt Lake City, UT 84112, USA}
\author{Zheng Liu}
\affiliation{Institute for Advanced Study, Tsinghua University, Beijing 100084, China}
\author{Miao Zhou}
\affiliation{Key Laboratory of Optoelectronic Technology and Systems of the Education Ministry of China, College of Optoelectronic Engineering, Chongqing University, Chongqing 400044, China}
\author{Xiaojuan Ni}
\affiliation{Department of Materials Science and Engineering, University of Utah, Salt Lake City, UT 84112, USA}
\author{Feng Liu}
\email[Corresponding author:]{fliu@eng.utah.edu}
\affiliation{Department of Materials Science and Engineering, University of Utah, Salt Lake City, UT 84112, USA}

\date{\today}

\begin{abstract}
The newly observed ($\sqrt{3}$$\times$$\sqrt{3}$) surface reconstruction in heteroepitaxial Si(111) thin films on metal substrates is widely considered as a promising platform to realize 2D Dirac and topological states, yet its formation mechanism and structural stability remain poorly understood, leading to the controversial terminology of ``multilayer silicene". Based on valence bond and conjugation theory, we propose a $\pi$-conjugation plus charge-transfer model to elucidate such a unique ``bamboo hat" surface geometry. The formation of planar ring-shaped $\pi$-conjugation and charge transfer from the rings to upper buckled Si atoms greatly lower the surface dangling bond energy. We justify this unconventional Si structural model by analyzing from first-principles surface stress tensors and surface energies as a function of strain. Within the same formalism, additional metastable surface reconstructions with the similar ``bamboo hat" features are predicted which opens possibilities to other exotic electronic states in Si.
\end{abstract}

\maketitle

$\pi$-conjugation has long been known to play a key role in stabilizing the carbon-based planar structures, such as benzene, graphite, and graphene \cite{1,2,3,4,5}. The other Group IV elements, however, have a much weaker tendency to form $\pi$-conjugation, because of their larger atomic radius. For example, all Si allotropes adopt a ``3D" bonding configuration with fully saturated covalent \textit{sp}$^3$ bonds. Weak $\pi$-conjugation has been found in Si(111)-(2$\times$1) surface within a linear chain structure \cite{6,7,8}, but the most typical hexagonal ring structure has never been seen. This underlies the difficulty in experimentally synthesizing the elusive freestanding form of silicene \cite{9,10,11,12,13}.

Interestingly, a ``planar" hexagonal ring-shaped structure has been observed in the surface of epitaxially grown Si(111)-($\sqrt{3}$$\times$$\sqrt{3}$) thin films \cite{14,15,16,17}, which is dubbed as ``multilayer silicene" by some researchers. However, previous first-principles calculations have invalidated a stacked silicene structure, which spontaneously transforms into the bulk \textit{sp}$^3$ structure with just two layers of stacking \cite{16,17}. Thus, the silicene-like electronic properties, such as Dirac cone, should only be attributed to the unique ($\sqrt{3}$$\times$$\sqrt{3}$) surface reconstruction. Understanding its formation mechanism will help resolve the long standing ``silicene" puzzle, shedding new light on understanding the difficulties of growing freestanding silicene.

On the other hand, the surface properties of Si have been extensively studied for many decades, because of its extraordinary importance to electronic devices \cite{18,19,20,21,22,23,24}. The basic surface reconstruction of Si was considered well understood, such as the (7$\times$7) reconstruction for the annealed and the (2$\times$1) reconstruction for the cleaved Si(111) surface. So the newly observed ($\sqrt{3}$$\times$$\sqrt{3}$) surface in heteroepitaxial Si(111) thin films is a big surprise, because it is fundamentally different from all the previous models, especially considering the unusual planar ring structure unexpected for Si. Clarifying the physical mechanism of such a unique surface reconstruction is thus of particular importance, which may profoundly renew our interest in Si surface and open a new route to realizing Dirac and topological bands in Si surface \cite{25,26,27,28}, as an interesting alternative to silicene.

In this Letter, we first revisit the traditional surface reconstructions of Si and then propose a $\pi$-conjugation plus charge-transfer model to explain the unexpected stability of the Si(111)-($\sqrt{3}$$\times$$\sqrt{3}$) surface reconstruction with a ``bamboo hat" bonding geometry. Based on density functional theory (DFT) calculations \cite{29}, we further evaluate the effect of strain on surface energies of both ($\sqrt{3}$$\times$$\sqrt{3}$) and (2$\times$1) superstructures to explain why this unusual reconstruction occurs when ultrathin Si(111) film is grown on substrates. Finally, we investigate the possible configurations of Si(111)-($\sqrt{21}$$\times$$\sqrt{21}$)  reconstruction as observed in a recent experiment \cite{30}, which can also be explained by the $\pi$-conjugation and charge-transfer model.

For the (111)-oriented Si film, different from the $\pi$-stacking interaction in graphite, Si prefers to form \textit{sp}$^3$ hybridized $\sigma$ bonds and a crossover between silicene and bulk Si is expected when the Si film is thicker than just two layers \cite{16,17}. The ($\sqrt{3}$$\times$$\sqrt{3}$) superstructure therefore represents a new surface reconstruction in the Si(111) surface, differing from (7$\times$7) and (2$\times$1) surface reconstruction of bulk-terminated Si. We note that there is also a monolayer ($\sqrt{3}$$\times$$\sqrt{3}$) silicene reported on the Ag and Ir substrate \cite{12,31}, which is controlled by the substrate induced strain \cite{32}. Although it does serve as the buffer layer for the subsequent ($\sqrt{3}$$\times$$\sqrt{3}$) reconstructed layers \cite{17,33}, its formation is clearly different.

\begin{figure}[tbp]
\includegraphics[scale=0.3]{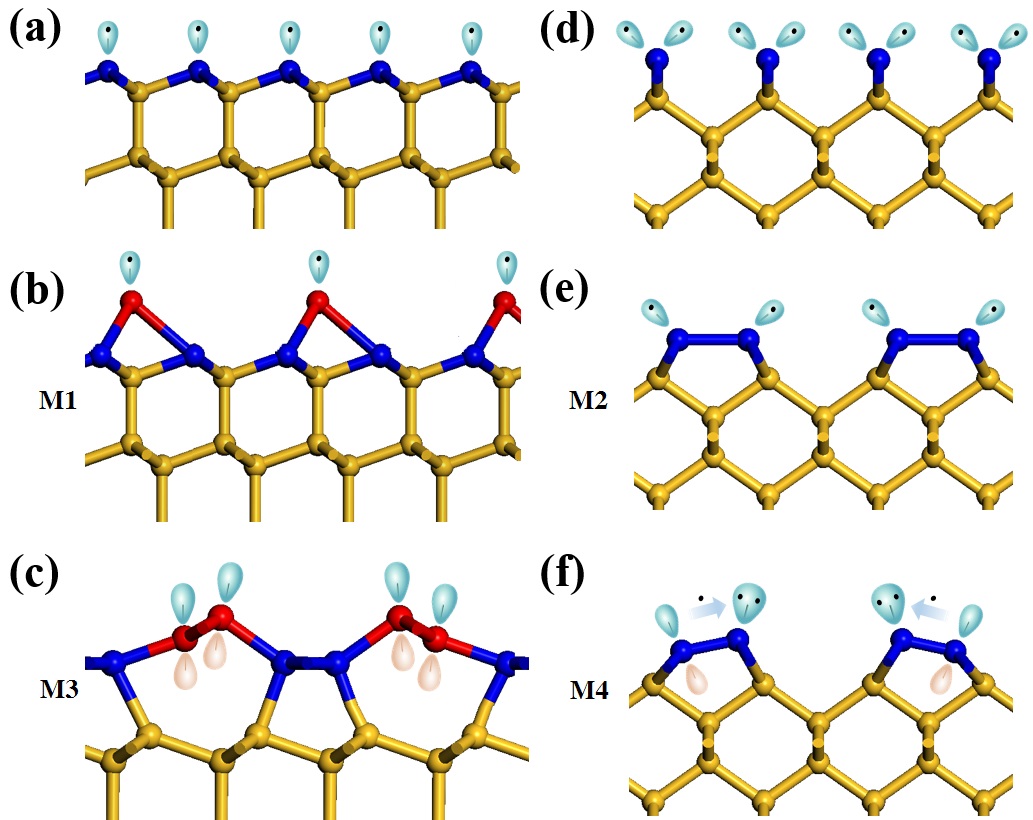}
\caption{Schematic view of different mechanisms of Si surface reconstructions. (a) and (d) Bulk terminated Si(111) and Si(100) surface with one and two dangling bonds, respectively. (b) The structure with the adatoms (red) that saturate the surface dangling bonds in Si(111)-(7$\times$7) surface. (c) Formation of $\pi$-conjugated chain (red) through couplings between neighboring \textit{p} orbitals in Si(111)-(2$\times$1) surface. (e) The formation of dimers that reduce one dangling bond per surface atom. (f) The buckling induces charge transfer from the down-buckled to up-buckled atom to form empty \textit{p}$_z$ and electron lone-pairs, respectively, in the Si(100)-(2$\times$1) surface. (color online)}
\label{f1}\end{figure}

In forming a Si surface, dangling bonds are created, e.g. one and two dangling bonds per Si atom in bulk-terminated Si(111) and Si(100) surface, respectively, as shown in Fig.~\ref{f1}a, d, which are highly unstable. To lower the high surface energy, the dominant mechanism is to lower the dangling-bond energy through surface reconstructions. In principle, there are two ways to lower the dangling-bond energy \cite{18,34,35,36}. One is obviously to remove the dangling bonds, which can be achieved by adsorbing adatoms directly over the surface layer as shown in Fig.~\ref{f1}b (named mechanism M1 thereafter) or creating dimers to form a covalent bond between two dangling bonds as shown in Fig.~\ref{f1}e (M2). The other way is to take the advantages of $\pi$-conjugation (M3) and charge-transfer (M4). The $\pi$-conjugation can be achieved through coupling between neighboring \textit{p} orbitals, in either linear-chain or ring shape in principle. So far, however, only the $\pi$-conjugated chain was reported on the Si(111)-(2$\times$1) surface (Fig.~\ref{f1}c). Charge-transfer, as manifested in dimer buckling (see Fig.~\ref{f1}f), lowers the energy by the \textit{Jahn}-\textit{Teller} effect, transferring electrons from the down-buckled to the up-buckled atom to form empty and filled (lone pair) dangling bonds, respectively.

In general, two or more mechanisms cooperate to stabilize a surface reconstruction. For example, in the Si(111) surface, the most stable (7$\times$7) superstructure was explained by dimer-adatom-stacking-fault model \cite{24}, which consists of both M1 and M2, and the metastable (2$\times$1) surface was clarified by a combination of M2 and M3 (Pandey model) \cite{8}. Also, M3 and M4 were used to explain the Si(100)-(2$\times$1) reconstruction \cite{37}. Here, we will apply some of these same principles to understand the recently observed Si(111)-($\sqrt{3}$$\times$$\sqrt{3}$) surface \cite{14,15,16,17}.

\begin{figure}[tbp]
\includegraphics[scale=0.25]{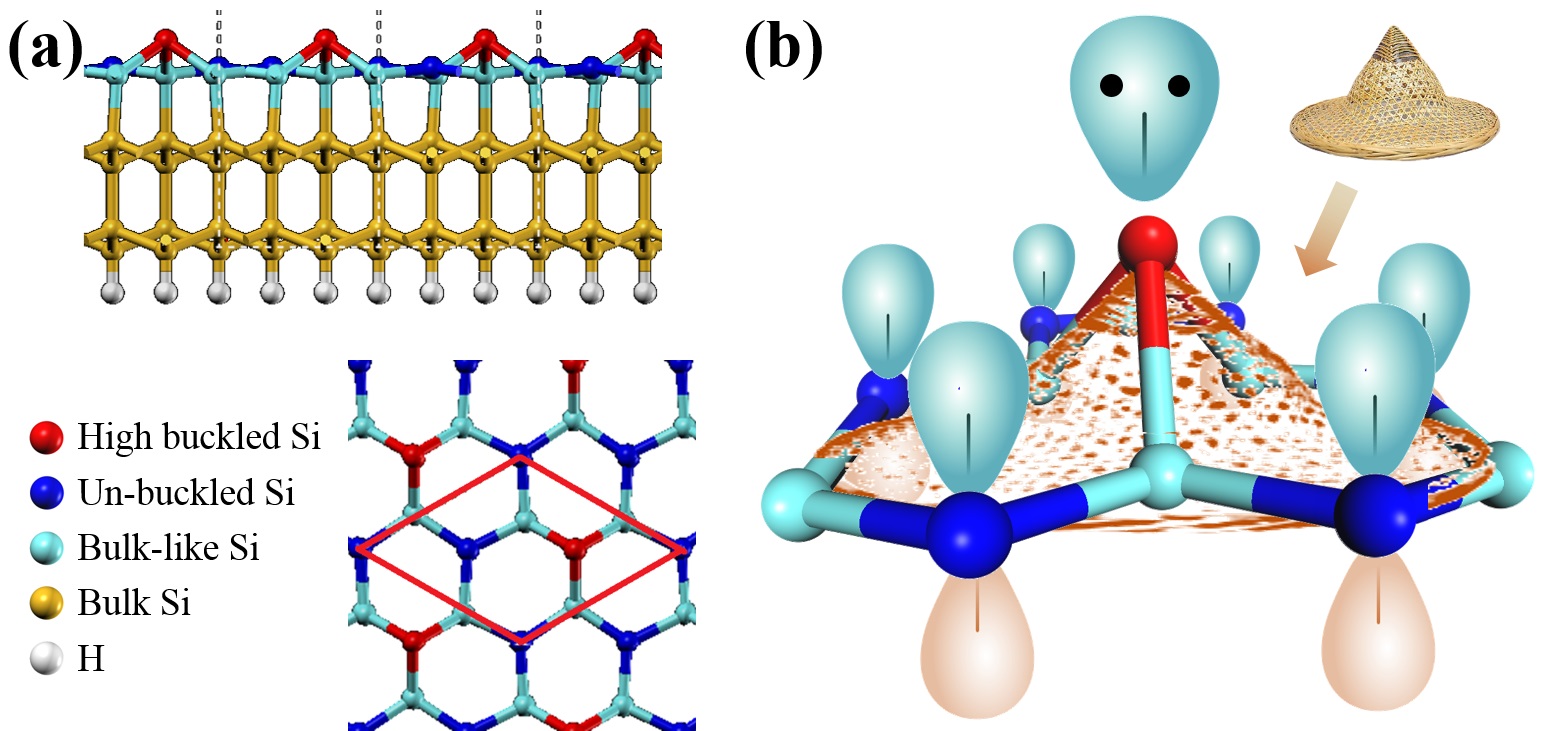}
\caption{Analysis of the structural property of the ($\sqrt{3}$$\times$$\sqrt{3}$) superstructure in Si(111) surface. (a) Side and top view of the ($\sqrt{3}$$\times$$\sqrt{3}$) surface. The red lines indicate the unit cell. (b) Schematic view of $\pi$-conjugation and electron lone pair in the ($\sqrt{3}$$\times$$\sqrt{3}$) surface with the inset indicating the bamboo hat geometry. (color online)}
\label{f2}\end{figure}

As shown in Fig.~\ref{f2}a, among six Si atoms in the ($\sqrt{3}$$\times$$\sqrt{3}$) surface, three atoms (light blue) are bonded with the underlying Si (BL-Si), one (red) is highly buckled (HB-Si), and the other two (deep blue) are almost unbuckled (UB-Si). The BL-Si atoms form four bonds with surface and underneath Si atoms showing an \textit{sp}$^2$ + $\sigma$ hybridization. In contrast, the UB-Si and HB-Si atoms have only three bonds with their nearest neighbors (NN) and form a ``planar" structure with a small buckling height around 0.2{\AA} and a typical tetrahedral structure, respectively. The former indicates an \textit{sp}$^2$ hybridized state with an un-hybridized \textit{p}$_{z}$ orbital and the latter has an \textit{sp}$^3$ hybridized state with a lone pair of electrons. Similar to $\pi$-conjugation in graphene, to increase the stability, the UB-Si with un-hybridized \textit{p}$_{z}$ orbitals, arranged in a hexagonal lattice, form delocalized $\pi$ bond through conjugated $\pi$-$\pi$ interaction. On the other hand, the HB-Si is further stabilized via charge transfer from the UB-Si to form a lone pair (Fig.~\ref{f2}b).

The overall structure of this peculiar bonding configuration resembles closely to a bamboo hat shape, as shown in Fig.~\ref{f2}b. In the following, we refer to it as BHS surface. Note that different from M4 where charge is transferred directly between two bonded atoms, the charge-transfer here occurs indirectly through the bridging BL-Si atoms; also different from M3 where $\pi$-conjugation is formed by the NNs, the $\pi$-conjugation in the BHS surface is in between the next NNs. To the best of our knowledge, this is the first example that has a planar ring-shaped $\pi$-conjugation for Si atoms.

\begin{figure}[tbp]
\includegraphics[scale=0.3]{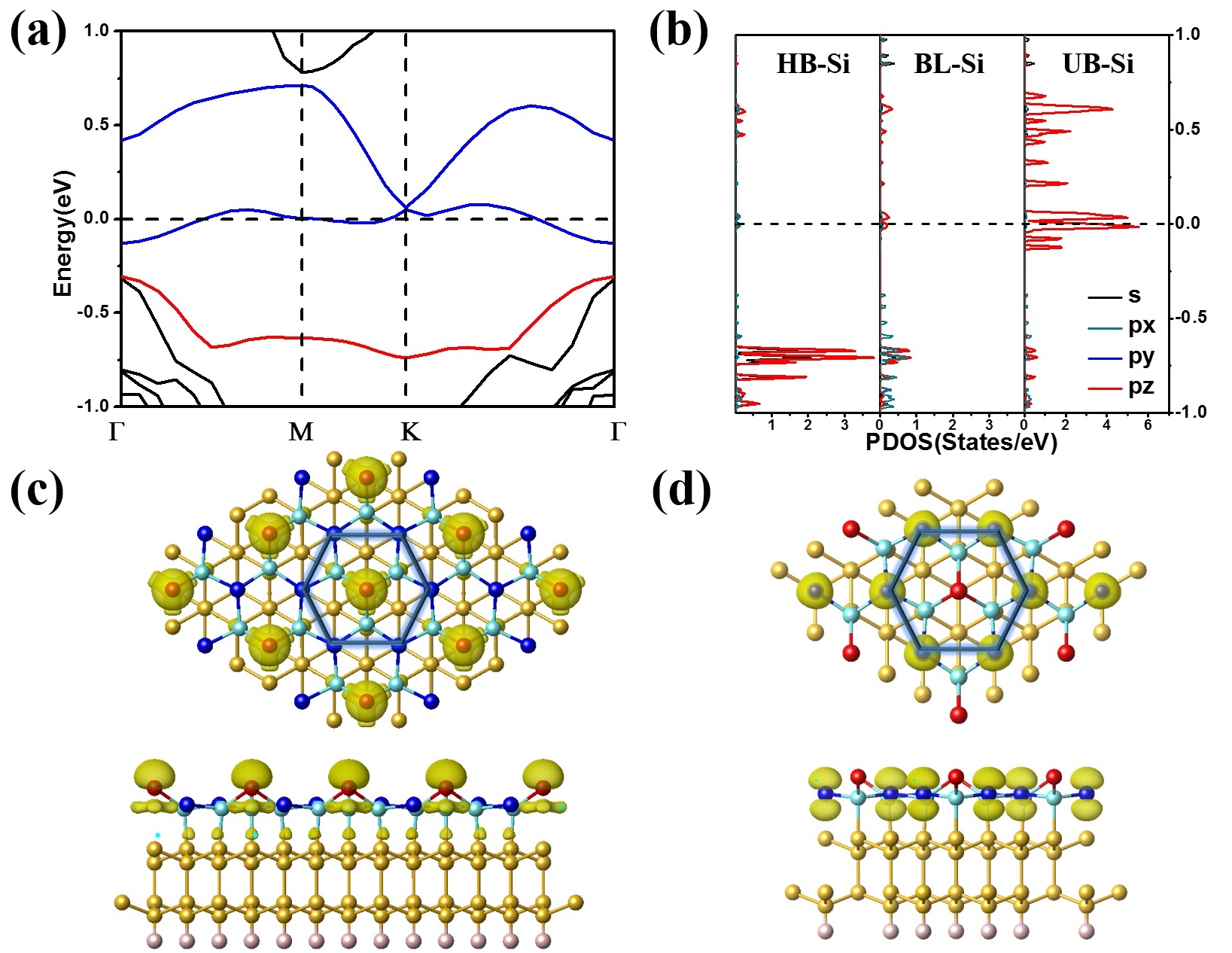}
\caption{ DFT simulation. (a) and (b) Band structure and the projected density of states of the ($\sqrt{3}$$\times$$\sqrt{3}$) surface reconstruction. The one-fourth filled un-hybridized \textit{p}$_z$ orbitals of the UB-Si form the Dirac bands (blue bands), while \textit{sp}$^3$ hybridized orbitals of HB-Si are fully occupied and lie under the Fermi level (red band). (c) and (d) Top and side view of the partial charge distribution of the FB and the Dirac band displaying a clear \textit{sp}$^3$ and \textit{sp}$^2$ hybridized orbital shape around HB-Si and UB-Si atoms, respectively. Central blue hexagon highlights the $\pi$-conjugation. (Color online)}
\label{f3}\end{figure}

To further verify this intriguing bonding structure for Si, we calculated the band structure along special \textit{K}-points and the projected density of states (PDOS) around the Fermi level for the BHS surface. As shown in Fig.~\ref{f3}a, b, the nearly flat band (FB) associated with the \textit{sp}$^3$ hybridized states for HB-Si atoms lies below Fermi level (red band) and is fully filled, indicating a lone pair. Similar to graphene, the hexagonal lattice consisting of un-hybridized \textit{p}$_{z}$ orbitals of the UB-Si produces a Dirac cone (blue bands in Fig.~\ref{f3}a), indicating the $\pi$-conjugation among UB-Si atoms. Different from the half-filled $\pi$ bands in graphene where the Fermi level is located exactly at the Dirac point, the Dirac bands here are one-fourth filled, which confirms the electron transfer from the UB-Si to the HB-Si atoms. Due to a longer hopping distance, the calculated band width and the Fermi velocity is relatively smaller than those in graphene. In addition, to confirm the hybridization nature of the surface atoms, we calculated the partial charge densities for the Dirac bands and the underneath FB, respectively, as shown in Fig.~\ref{f3}c, d. The charge densities for the FB are mainly localized around the HB-Si atoms, and a clear \textit{sp}$^3$ hybridized orbital shape can be seen (Fig.~\ref{f3}c). On the other hand, the Dirac bands are verified to consist of the dumbbell-shaped \textit{p}$_{z}$ orbitals from the UB-Si atoms arranged in the hexagonal lattice (blue hexagon in Fig.~\ref{f3}d). To better understand the surface electronic structure, we also calculated a ($\sqrt{3}$$\times$$\sqrt{3}$) monolayer structure model and constructed an effective tight-binding Hamiltonian \cite{29}, which confirm that the BHS surface is stabilized by the cooperative effects of ring-shaped $\pi$-conjugation and charge-transfer effect.

Next, we analyze why the BHS surface is observed rather than the bulk-terminated (2$\times$1) surface reconstruction when Si is grown on various substrates epitaxially. Because different growing methods can generate dramatically different strain/stress in the Si surface layers \cite{38,39}, we studied the effect of strain on these two surface reconstructions. For both systems, we calculated surface energy ($\gamma$), surface stress tensors ($\sigma$), and stress anisotropies (F) in the unstrained Si(111) surface using the following equations,
\begin{subequations}
    \begin{align}
        \gamma = \frac{1}{2A}(E^N_{slab} -N\Delta E)\label{eq1a}\\
        \Delta E = \frac{(E^N_{slab} -E^{N-2}_{slab})}{2}\label{eq1b}
    \end{align}
\end{subequations}
where A is the surface area, $E_{slab}^N$ is the total energy of slabs that contain N number of atomic layers, and $\Delta$E represents the total energy for one layer of bulk Si atoms, as calculated using Eqs.~\ref{eq1b}. To describe the surface stress tensor, for the (2$\times$1) superstructure, the \textit{x} and \textit{y} directions were set perpendicular to and along the $\pi$ conjugated chains, respectively, while for the BHS surface, the \textit{x} and \textit{y} directions were set arbitrarily given its structural isotropy (Fig. S3) \cite{29}. We used positive value to indicate tensile stress, and summarized the calculated results in Table I. The unstrained surface energies for ($\sqrt{3}$$\times$$\sqrt{3}$) and (2$\times$1) superstructures are 90.5 and 86.8 meV/{\AA}$^2$, respectively, which are relatively higher than the experimental value for the most stable Si(111) surface (76.8 meV/{\AA}$^2$ \cite{40}). Both systems exhibit a tensile stress with the (2$\times$1) surface having dramatic higher values than the BHS surface. Moreover, the (2$\times$1) surface shows a large surface stress anisotropy, caused by the alternating buckled $\pi$-bonded chains that tend to shrink the surface in the direction perpendicular to the chain.

\begin{table*}[tbp]
\label{tb1}
\caption{Surface energies ($\gamma$), stress tensors ($\sigma$), and stress anisotropies (\textit{F}) of ($\sqrt{3}$$\times$$\sqrt{3}$) and (2$\times$1) reconstructions in Si(111) surface calculated using the first-principles methods.  }
\begin{tabular}{ccccccccccc}
\hline\hline
\multirow{2}{*}{No. of Layers} & \multicolumn{4}{c}{$\sqrt{3}$$\times$$\sqrt{3}$} & \multicolumn{4}{c}{2$\times$1}\\
   & $\gamma(meV/{\AA}^2)$ & $\sigma_{xx}(meV/{\AA}^2)$ &$\sigma_{yy}(meV/{\AA}^2)$ & $\textit{F}(meV/{\AA}^2)$ & $\gamma(meV/{\AA}^2)$ & $\sigma_{xx}(meV/{\AA}^2)$ &$\sigma_{yy}(meV/{\AA}^2)$ & $\textit{F}(meV/{\AA}^2)$ \\
 \hline
3   &86.7   &45.6   &45.6   &0.0    &89.9   &268.4   &80.5  &187.9\\
4	&89.9	&53.7   &58.0	&-4.3	&85.5   &250.3   &81.1  &169.2\\
5	&89.9	&54.3   &54.9   &-0.6   &86.8   &269.0   &89.3	&179.7\\
6	&90.5	&53.7   &53.7   &0.0	&86.8	&271.5   &89.9	&181.6\\
7	&90.5	&51.2   &53.7   &-2.5	&86.8	&270.3 	 &91.1	&179.2\\
8	&89.9	&49.3   &52.4   &-3.1	&86.8   &269.0 	 &86.8	&182.2\\
 \hline\hline
\end{tabular}%
\end{table*}

Using the above calculation results, we can estimate the relative surface energy under strain using: $\gamma_s(\varepsilon)=\gamma_0 + \sigma\cdot\varepsilon$, where $\gamma_0$ is the calculated unstrained surface energy, $\sigma$ is the stress tensor, and $\varepsilon$ is the strain with positive value indicating tensile strain. Surface energies as a function of strain are (see also Fig. S4)
\begin{subequations}
    \begin{align}
        \gamma_{(\sqrt{3}\times\sqrt{3})}(\varepsilon) = 90.5 + 106.7\varepsilon\label{eq2a}\\
        \gamma_{(2\times1)}(\varepsilon) = 86.8 + 359.5\varepsilon\label{eq2b}
    \end{align}
\end{subequations}
For the unstrained Si(111) surface, the surface energy of the (2$\times$1) reconstruction is about 3.7 meV/{\AA}$^2$ lower than the BHS surface. However, the (2$\times$1) surface is more sensitive to strain than the BHS surface because of its higher surface stress. Consequently, the BHS surface becomes more stable than the (2$\times$1) surface when the applied tensile strain $\varepsilon$ is higher than 1.5\%, as also indicated by the black arrow in Fig. S4. This explains why the ($\sqrt{3}$$\times$$\sqrt{3}$) reconstruction was seen on the Ag(111) substrate \cite{17,32}, as the Ag(111) in-plane lattice constant is larger than that of Si. It is worth noting that surface stress is essentially created by electron redistribution around the surface atoms, so the charge transfer from the Ag substrate to the grown Si multilayers may affect the surface stress. Therefore, we studied this possible effect by calculating the charge differential for different layers of Si on Ag, which shows that the charge transfer mainly occurs around the area where Ag and Si atoms are in direct contact, and the effect to the surface layer becomes negligible when the Si layer is thicker than two. We also note that since the ground state Si(111)-(7$\times$7) reconstruction for thicker Si layers involves multilayer reconstruction of Si atoms, it must give the way to the BHS surface at the early stage of epitaxial growth.

\begin{figure}[tbp]
\includegraphics[scale=0.3]{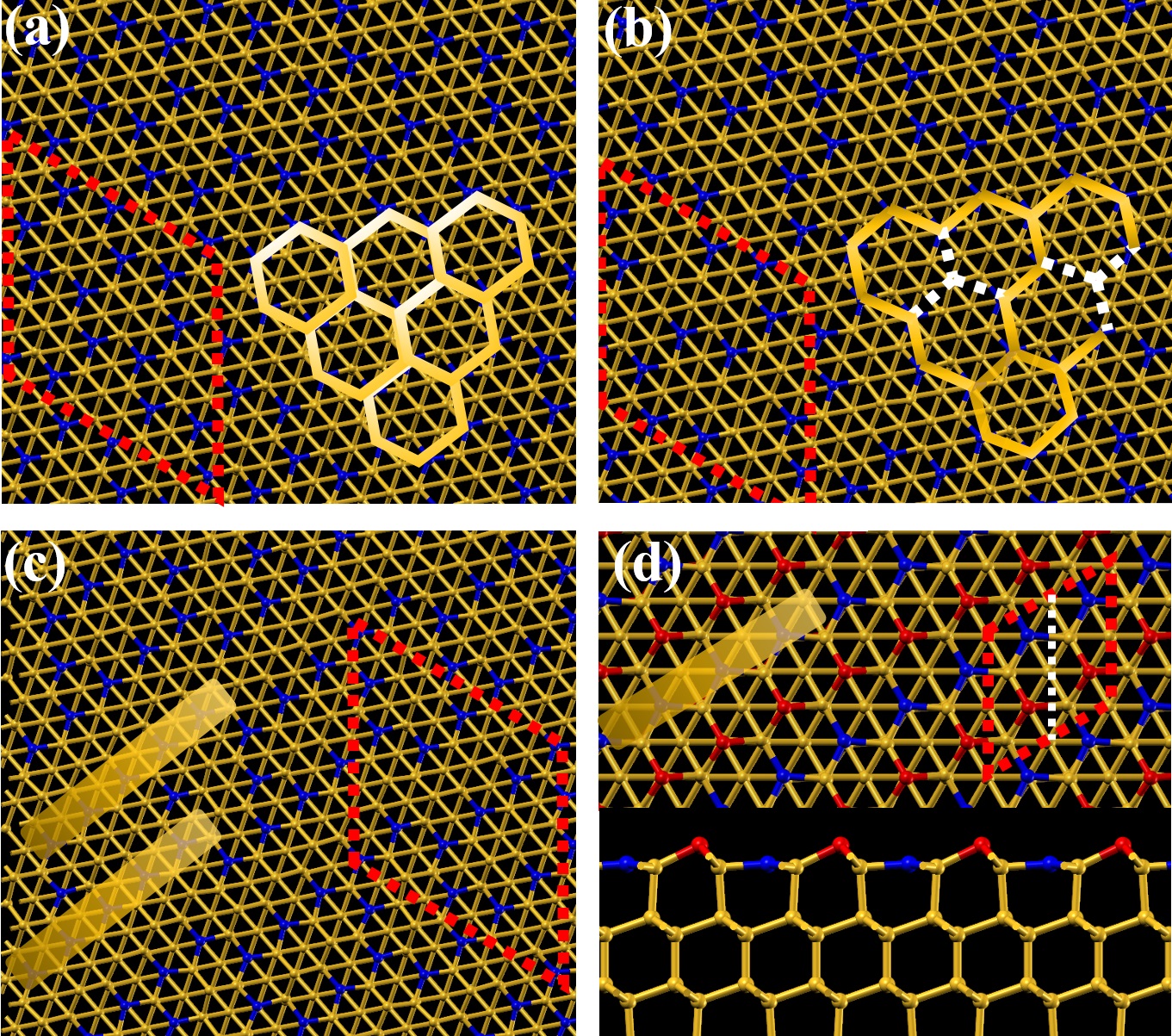}
\caption{$\pi$-conjugation and charge-transfer model in the ($\sqrt{21}$$\times$$\sqrt{21}$) unit cell, as indicated by the red dashed rhombus. (a) Reproduced ($\sqrt{3}$$\times$$\sqrt{3}$) surface reconstruction with $\pi$-conjugation formed by the unbuckled Si atoms (blue atoms), as indicated by yellow super-hexagons. (b) Metastable high symmetric ($\sqrt{21}$$\times$$\sqrt{21}$) surface reconstruction. The white dashed lines indicate the breaking of the ($\sqrt{3}$$\times$$\sqrt{3}$) hexagonal $\pi$-conjugation. (c) Linear $\pi$-conjugation surface reconstruction as indicated by yellow ribbons. (d) Ideal buckling model for the (2$\times$1) surface reconstruction.(Color online) }
\label{f5}\end{figure}

Finally, we extended our calculation from ($\sqrt{3}$$\times$$\sqrt{3}$) to ($\sqrt{21}$$\times$$\sqrt{21}$) superstructure to explore other possible buckled surface geometries that may be observed during the ``silicene" growing processes \cite{29}. We tested thirty different initial surface configurations and all the relaxed structures showed a buckled surface geometry, where half surface Si atoms (21/42 atoms) bond with the underlying Si layer. Among the other half Si atoms with dangling bonds, several Si atoms exhibit an \textit{sp}$^2$ hybridization with nearly flat geometry, while the others buckled up (HB-Si) with an average buckling height around 1.1{\AA} (1.088-1.142{\AA}) showing an \textit{sp}$^3$ hybridization with electron lone pair, (see Fig. S5). This indicates charge transfer from the nearly-flat to the HB-Si atoms \cite{29}. As listed in Table S I, most of the structures have eight or nine HB-Si atoms except two structures that contain seven and ten HB-Si atoms, respectively \cite{29}. As shown in Fig.~\ref{f5}a, the structure with seven HB-Si atoms is actually a ``local" BHS surface. More importantly, the HB-Si atoms of all the structures show mainly two patterns as demonstrated by hexagons and ribbons in Fig. S6 \cite{29}, indicating the existence of either hexagonally or linearly arranged $\pi$-conjugations formed by the un-hybridized \textit{p}$_z$ orbitals of the unbuckled Si atoms (Fig.~\ref{f5}). This further confirms the $\pi$-conjugation plus charge-transfer mechanism in the compressively strained Si(111) surface to be general.

The structure with linear $\pi$-conjugation was found having the lowest energy (Fig.~\ref{f5}c). We noticed that this linearly conjugated pattern is similar to the buckling model (BM) proposed for the (2$\times$1) surface reconstruction (Fig.~\ref{f5}d) \cite{41}, so we also calculated the surface energy of the BM. Although the BM surface has almost the same unstrained surface energy as the BHS surface, it is easily relaxed to the more stable Pandy $\pi$-bonded chain structure due to their similar strong anisotropic stress feature, indicating its metastability. Interestingly, a new high symmetric ($\sqrt{21}$$\times$$\sqrt{21}$) superstructure with isotropic stress was discovered with nine HB-Si atoms (Fig.~\ref{f5}b), which highly resembles a recently experimentally observed structure in Si(111) surface \cite{30}. Besides the seven HB-Si atoms that form the ($\sqrt{3}$$\times$$\sqrt{3}$) triangular lattice, two more Si atoms sitting at the center of the two green triangles are slightly buckled up (see Fig. S5) \cite{29}. The buckling of these two atoms partially breaks the ($\sqrt{3}$$\times$$\sqrt{3}$) hexagonal $\pi$-conjugation (dashed white lines in Fig.~\ref{f5}b), which increases slightly the energy by 0.03eV of the whole system compared to the ``local" BHS surface. The surface energies as a function of strain were also calculated to be: $\gamma_{(\sqrt{21}\times\sqrt{21})}(\varepsilon)=91.1+136.1\varepsilon$ (meV/{\AA}$^2$), which is very close to the BHS surface, indicating high structural stability under compressive strain.

In conclusion, we have revealed a cooperative mechanism of hexagonal ring-shaped $\pi$-conjugation and charge-transfer, which stabilizes the newly observed epitaxial Si(111)-($\sqrt{3}$$\times$$\sqrt{3}$) surface with a bamboo hat bonding geometry. It differs dramatically from the commonly known bonding structures in Si surface. The reason that ($\sqrt{3}$$\times$$\sqrt{3}$) structure is observed on substrate rather than the bulk-terminated Si(111)-(2$\times$1) surface can be explained by their different response to external strain induced by lattice mismatch. These findings broaden our knowledge of reconstruction mechanisms in Si surface, which also shed new light on understanding the difficulties of growing monolayer silicene. It may also have important implication in other epitaxial semiconductor films when they are grown on strained substrate.

W. J. was supported by the National Science Foundation-Material Research Science \& Engineering Center (NSF-MRSEC grand No. DMR-1121252). F. L. acknowledges additional support from U.S. DOE-BES (Grant No. DE-FG02-04ER46148). We thank the CHPC at the University of Utah and DOE-NERSC for providing the computing resources.

\end{document}